\documentclass[preprint]{aastex}

\makeatletter

\begin{document}
\title{Magnetic dynamo action in random flows with zero and finite correlation times}
\author{Joanne Mason} 
\author{Leonid Malyshkin}
\affil{Department of Astronomy and Astrophysics, University of Chicago, 
5640 S. Ellis Ave, Chicago, IL 60637; {\sf jmason@flash.uchicago.edu; leonmal@flash.uchicago.edu}}
\author{Stanislav Boldyrev}
\affil{Department of Physics, University of Wisconsin at Madison, 1150 University Ave, 
Madison, WI 53706; {\sf boldyrev@wisc.edu}}
\author{Fausto Cattaneo}
\affil{Department of Astronomy and Astrophysics, University of Chicago, 
5640 S. Ellis Ave, Chicago, IL 60637; {\sf cattaneo@flash.uchicago.edu}}
\date{\today}

\begin{abstract}
Hydromagnetic dynamo theory provides the prevailing theoretical description for the origin of magnetic fields in the universe. Here we consider the problem of kinematic, small-scale dynamo action driven by a random, incompressible, non-helical, homogeneous and isotropic flow. In the Kazantsev dynamo model the statistics of the driving flow are assumed to be instantaneously correlated in time. Here we compare the results of the model with the dynamo properties of a simulated flow that has equivalent spatial characteristics as the Kazantsev flow but different temporal statistics. In particular, the simulated flow is a solution of the forced Navier-Stokes equations and hence has a finite correlation time. We find that the Kazantsev model typically predicts a larger magnetic growth rate and a magnetic spectrum that peaks at smaller scales. However, we show that by filtering the diffusivity spectrum at small scales it is possible to bring the growth rates into agreement and simultaneously align the magnetic spectra.
\end{abstract}


\maketitle

\section{Introduction}

Magnetic fields are thought to be present in a variety of astrophysical objects, including stars, planets, accretion 
discs, galaxies and the interstellar and intergalactic media. In some cases the magnetic field has been directly 
detected and its properties and consequences have long been known (e.g. the Sun's large-scale magnetic field that is responsible for the 22-year sunspot cycle), while the active role of 
magnetic fields in other astrophysical phenomena has only recently been recognised (such as its role in angular 
momentum transport in accretion discs). While in some cases it is possible that the magnetic field could be the remnant of a 
decaying fossil field, in others it is believed that the field is sustained against the action of Ohmic dissipation via 
the operation of a hydromagnetic dynamo -- a mechanism through which the kinetic energy of the plasma is 
converted into magnetic energy. Understanding the evolution of magnetic fields in the universe remains one of the most
important unsolved problems in astrophysics.

Due to the enormous scale of astrophysical objects, in virtually all cases it is believed that the flow that leads to dynamo action is turbulent. The 
generated magnetic field then exhibits complex spatial and temporal characteristics on a wide range of scales, a complete understanding of which represents a formidable problem. Progress with astrophysical dynamo theory is therefore typically made by breaking 
the problem down into a number of fundamental sub-problems. A natural starting point is to consider the initial stages 
of evolution of a weak seed field. During this kinematic phase the field is assumed to be so weak as to have no 
dynamical effects on the turbulence. The Navier-Stokes equations governing the evolution of the flow then 
decouple from the induction equation governing the magnetic evolution, viz.
\begin{equation}
\label{eq:induction}
\frac{\partial \mathbf{B}}{\partial t}=\nabla \times (\mathbf{u} \times \mathbf{B})+\eta \nabla^2 \mathbf{B}, \quad \nabla \cdot \mathbf{B}=0
\end{equation}
where $\eta$ is the magnetic 
diffusivity. The problem therefore reduces considerably to solving equations (\ref{eq:induction}), given a flow $\mathbf{u}(\mathbf{x},t)$. 
If the flow is a kinematic dynamo the magnetic field grows exponentially and the task is  to 
determine the magnetic growth rate and the preferred lengthscale of the dynamo instability.  In reality, the
exponential growth cannot continue indefinitely. Eventually 
the generated magnetic field will be strong enough to react back on the flow (via the Lorentz force in the momentum equation) and the growth will 
saturate. In this nonlinear phase the aim is to understand the dynamics of the saturation process and to 
determine the amplitude of the resulting field. It is important to note a further distinction that is often made: Large
scale dynamos are those in which the magnetic
field has a scale of variation much larger than the characteristic velocity
correlation length, while in small scale dynamos the magnetic field varies on scales comparable
with or smaller than the scale of the driving flow. 

In this paper we shall be concerned with the kinematic, small-scale dynamo problem. There are essentially three different avenues of research that have developed for addressing this scenario. They differ in the way in which the flow is specified. In the first, one considers the amplification of magnetic fields in a laminar flow that has a given analytical form (see the monograph by \cite{childressg1995}). The second method is similar, except that the flow is specified numerically. It is typically either a solution of the Navier-Stokes equations or it is synthesized to have certain properties. In the third approach, the flow is assumed to be a random process with prescribed statistics. In this case the aim is then to determine the corresponding statistical properties of the magnetic field. The work presented in this paper concerns a comparison of the second and third approaches.

The third avenue of research mentioned above (that of considering field amplification in temporally random velocity fields with given statistical properties) was pioneered by  \cite{kazantsev1968}, who considered the case of a Gaussian distributed random flow with zero mean that is instantaneously correlated in time and homogeneous and isotropic in space. In this case it  follows that the flow is completely specified by the two-point, two-time velocity covariance tensor 
\begin{equation}
\label{eq:corr_v}
\langle u_i(\mathbf{x},t) u_j(\mathbf{x+r},t+\tau) \rangle = \kappa_{ij}(\mathbf r) \delta (\tau),
\end{equation}
where $\langle ... \rangle$ denotes an ensemble average and $\delta(\tau)$ is the Dirac delta function. We note that such a statistical description is also often referred to as a Kraichnan statistical ensemble, in reference to \cite{kraichnan1968} who studied the evolution of a passive scalar field in a turbulent flow with the same statistical properties. The Gaussian property implies that all odd moments 
vanish and all even 
moments can be expressed in terms of products of the above second order moments. The most general form of 
the isotropic tensor $\kappa_{ij}(\mathbf{r})$ is (see, e.g., \cite{batchelor1953})
\begin{equation}
\label{eq:kappa_ij}
\kappa_{ij}(\mathbf r)=\kappa_N(r) \left( \delta_{ij}-\frac{r_i r_j}{r^2} \right)+\kappa_L(r) \frac{r_i r_j}{r^2} 
+g(r) \epsilon_{ijk}r_k,
\end{equation}
where $\delta_{ij}$ is the unit diagonal tensor, $\epsilon_{ijk}$ is the unit alternating tensor, $r=|\mathbf{r}|$ and 
summation over repeated indices is assumed. For an incompressible flow it follows that 
$\kappa_N=\kappa_L+(r/2)(d\kappa_L/dr)$. 

Assuming that the magnetic field is also homogeneous and isotropic, we have 
\begin{equation}
\label{eq:corr_b}
\langle B_i(\mathbf{x},t) B_j(\mathbf{x}+\mathbf{r},t) \rangle=M_N(r,t) \left( \delta_{ij}-\frac{r_i r_j}{r^2} 
\right)+M_L(r,t) \frac{r_i r_j}{r^2} +K(r,t) \epsilon_{ijk}r_k,
\end{equation}
with the solenoidal constraint implying $M_N=M_L+(r/2) (dM_L/dr)$. Interestingly, it is then possible to derive a closed system of 
equations for the evolution of the magnetic correlation functions ($M_L$ and $K$) in terms of the corresponding 
functions for the velocity ($\kappa_L$ and $g$) \citep{kazantsev1968, vainshteink1986}. Recently, \cite{boldyrevetal2005} established that the 
equations have a self-adjoint structure, implying that the eigenvalues of the system are real.
In the absence of kinetic helicity ($g= 0$) the Kazantsev model predicts zero magnetic helicity ($K=0$) and the equation for $M_L$ reads
\begin{equation}
\label{eq:M_L}
\frac{\partial M_L}{\partial t} =
\kappa \frac{\partial^2 M_L}{\partial r^2}+\left(\kappa'+\frac{4}{r} \kappa \right) \frac{\partial  M_L}{\partial 
r}+\left(\kappa''+\frac{4}{r}\kappa' \right)M_L
\end{equation}
where $\kappa(r)=\kappa_L(0)-\kappa_L(r)+2\eta$.
Formally, the eigenvalue problem for $M_L(r,t)=M(r)\exp(\lambda t)$ is to be solved in $r\in[0,\infty]$.  In the non-helical case (the case to be considered here) it can be shown that the eigenvalues form a discrete spectrum. The eigenmodes are spatially bound and satisfy the boundary conditions $M(0)=const.$ (where for the kinematic dynamo problem the constant is arbitrary and is related to the value of the seed field) and $M(r)$ decays exponentially to zero as $r \rightarrow \infty$ (see, e.g., \cite{malyshkinb2007}). 

The Kazantsev dynamo model is a very useful tool. It reduces the problem of solving a partial 
differential equation for the evolution of the three-dimensional vector field $\mathbf{B}(\mathbf{x},t)$ 
into two ordinary differential equations for the evolution of the scalars $M_L(r,t)$
and $K(r,t)$, given the corresponding quantities for the flow. In the former case, computational power limits 
studies to moderately turbulent systems with the Reynolds numbers $R_e=ul/\nu$ (where $\nu$ is the viscosity) and 
$R_m$ being 
approximately equal and of the order of a few thousand. The validity of extrapolating the 
results to astrophysical situations where the magnetic Prandtl number ($P_m=R_m/R_e$) is typically either tiny
(e.g., in the solar convection zone or the Earth's core) or huge (e.g., in the ISM) and the Reynolds numbers are enormous is unknown.  By contrast, the Kazantsev equations 
can be solved relatively 
inexpensively and Reynolds numbers up to $10^{12}$ are permissible (the limit being only that of the double precision accuracy of the standard computer floating-point format (see \cite{malyshkinb2008}).

The Kazantsev model is the only available analytical model for dynamo action. It has therefore become very popular and over the years the theory has been built upon considerably. For example, \cite{kulsruda1992} developed a detailed theory for the spectra of magnetic fluctuations driven by an incompressible velocity field with a Kolmogorov spectrum $E(k) \sim k^{-5/3}$ (see also \cite{kraichnann1967}). The spectral theory was extended to describe $d$-dimensional flows with arbitrary degrees of compressibility by \cite{schekochihinbk2002}. The Kazantsev model has been solved for various different forms of the velocity correlator (see, e.g., \cite{zeldovichrs1990} and references therein) and the theory has been broadened to address the role of helicity (see, e.g., \cite{vainshteink1986, bergerr1995, kimh1997, boldyrevetal2005}), flows with varying degrees of roughness (e.g. \cite{boldyrevc2004, malyshkinb2010}), nonlinear effects \citep{kim1999} and the consequences on dynamo action of finite time correlated flows (e.g.~\cite{chandran1997, rogachevskiik1997, schekochihink2001}).

Ultimately, as with any simplified model, one must assess the extent to which 
the solutions of the model depend on the assumptions made, in this case regarding the statistics of the velocity. In particular, it is important to determine how accurately the Kazantsev dynamo model describes the situation in which the velocity is physically realistic, i.e.~when the flow satisfies the equations of electrically conducting fluid dynamics. In this first paper we accept the conditions of incompressibility, homogeneity and isotropy, and we also note that the assumption of Gaussianity is not believed to be of crucial importance. Indeed, direct numerical simulations have shown that a 
purely Gaussian velocity field amplifies magnetic fluctuations in a similar manner to a Navier-Stokes velocity (see \cite{tsinoberg2003}). However, we believe that it is important to note that in addition to the magnetic diffusivity, the only inputs to the Kazantsev model are the 
spatial correlation functions ($\kappa_L$ and $g$) of the velocity. All of the temporal properties of the flow are encapsulated by the $\delta$-function in 
equation
(\ref{eq:corr_v}). Indeed, the fact that the flow is instantaneously correlated in time is essential 
for the derivation of the 
closed system of equations governing the evolution of $M_L$ and $K$ (i.e. equation (\ref{eq:M_L}) in the 
non-helical case). If the advecting flow is finite-time 
correlated, as it will be in reality, these equations cannot be obtained in closed form.

Physically, one might expect that the results of the $\delta$-correlated model would smoothly match onto the 
short-time correlated case if the velocity correlation time is much smaller than the characteristic time for 
dynamo action. However, the 
question remains as to what happens as the correlation time of the flow increases. Indeed, it has recently been 
demonstrated that for the quasi two-dimensional dynamo problem, two flows with the same velocity 
spectrum but differing phase properties can have very different dynamo properties \citep{tobiasc2008a}. The goal 
of the present work is therefore to compare the characteristics of magnetic field generation in the instantaneously correlated Kazantsev flow with the dynamo properties of flows that have a similar spatial structure but that are solutions of the forced, incompressible, Navier-Stokes equation and hence have finite correlation times. 

Before proceeding, we would like to note that in the remainder of the paper it will sometimes prove useful to consider the correlators of the Fourier coefficients of the velocity, rather than the spatial correlators (\ref{eq:corr_v}) and (\ref{eq:kappa_ij}). A further advantage is that the physical 
meaning of the functions $\kappa_L(r)$ and $g(r)$ then becomes apparent. We therefore introduce the Fourier 
transform of the velocity
\begin{eqnarray}
\label{eq:uk}
\mathbf{u}(\mathbf{x},t) \equiv \int \tilde \mathbf{u}(\mathbf{k},t) \exp(i \mathbf{k}\cdot \mathbf{x}) 
d\mathbf{k}.
\end{eqnarray}
In the Kazantsev model, the velocity correlator in $\mathbf{k}$-space then takes the form (assuming incompressibility)
\begin{equation}
\label{eq:corr_v_k}
\langle \tilde u_l^*(\mathbf{k},t) \tilde u_m(\mathbf{k'},t+\tau) \rangle = \left\{ F(k) \left( \delta_{lm}-\frac{k_l k_m}{k^2} \right)+i \epsilon_{lmn} k_n G(k) \right\}
\delta(\mathbf{k'}-\mathbf{k}) 
\delta (\tau)
\end{equation}
where $^*$ denotes the complex conjugate.
The functions $\kappa_L(r)$ and
$g(r)$ can be obtained from $F(k)$ and $G(k)$ through three-dimensional Fourier transforms (see \cite{moniny1971}). In particular, we will use the relation 
\begin{equation}
\label{eq:kappa_L_F}
\kappa_L(r)=8 \pi \int_{0}^{\infty} \frac{[\sin(kr)-kr\cos(kr)]}{kr^3}F(k) dk
\end{equation}
We note that $\langle \mathbf{u}(\mathbf{x},t) \cdot \mathbf{u}(\mathbf{x},t+\tau)\rangle=3 \delta(\tau) 
\kappa_L(0)=2 \delta(\tau) \int F(k) d\mathbf{k}=2 \delta(\tau) \int 4\pi k^2 F(k)dk$. Similarly 
$\langle \mathbf{u}(\mathbf{x},t)\cdot (\nabla \times  \mathbf{u}(\mathbf{x},t+\tau)) \rangle=-2 \delta(\tau) \int 4 
\pi k^4 G(k)dk$. Thus the functions $F(k)$ and $G(k)$ in equation (\ref{eq:corr_v_k}) are related to the kinetic energy and helicity in the Kazantsev model. The equivalent magnetic correlator in Fourier space is
\begin{equation}
\label{eq:corr_B_k}
\langle \tilde B_l^*(\mathbf{k},t) \tilde B_m(\mathbf{k},t) \rangle = F_B(k,t) \left( \delta_{lm}-\frac{k_l 
k_m}{k^2} \right)+i \epsilon_{lmn} k_n G_B(k,t),
\end{equation}
where again $M_L(r,t)$ and $K(r,t)$ are related to $F_B(k,t)$ and $G_B(k,t)$ through three-dimensional Fourier 
transforms. In particular, $E_B(k,t)=4\pi k^2 F_B(k,t)$ and $H_B(k,t)=-8 \pi k^4 G_B(k,t)$ are the spectrum 
functions of magnetic energy and electric current helicity, respectively.

\section{Formulation of the problem}

We shall first solve the induction equation (\ref{eq:induction}) with a prescribed flow 
that is a statistically steady solution of the randomly driven Navier-Stokes equations 
\begin{equation}
\label{eq:momentum}
\frac{\partial \mathbf{u}}{\partial t}+(\mathbf{u} \cdot \nabla)\mathbf{u}=-\nabla p+\nu \nabla^2 
\mathbf{u}+\mathbf{f}, \qquad \nabla \cdot \mathbf{u}=0.
\end{equation}
Here $p$ is the pressure (whose role is to maintain incompressibility), $\nu$ is the viscosity and $\mathbf{f}$ 
is a random force, the specific properties of which we detail below. We have neglected the 
Lorentz force since we are considering the kinematic case in which the magnetic field is weak and hence doesn't 
affect the flow. The equations (\ref{eq:induction},\ref{eq:momentum}) are solved on the triply 
periodic domain $\mathbf{x}\in [0,2\pi]$. We 
note that although the Kazanstev model is defined on the infinite domain 
$r\in [0,\infty]$, it is anticipated (and verified below) that the periodic boundary conditions do not play a crucial role if it is ensured that 
the energy containing scales of the flow and the magnetic field are significantly smaller than $2\pi$. It is pointed out that this requirement, together with the restrictions placed by currently available 
computational power, does however severely limit the extent of the inertial range of the simulated flow.

In order to conduct a meaningful comparison with the Kazantsev model the simulated velocity must have the 
required 
spatial properties, i.e. it must be small-scale (significantly smaller than the box size $2\pi$) and spatially homogeneous and isotropic. For this first study we also 
restrict our attention to the non-helical case. The idea is to impose on the forcing function the properties that are required of the velocity. Then, at least in the strongly diffusive case where the solution of equation (\ref{eq:momentum}) represents a 
balance between the diffusive terms and the force, the flow will inherit those characteristics. We therefore choose 
a random, divergence-free, non-helical, homogeneous and isotropic force with Fourier coefficients 
\begin{equation}
\label{eq:force}
\tilde \mathbf{f}(\mathbf{k},t)=A a_\mathbf{k} (\mathbf{k}\times \mathbf{\hat e_k})\exp(i\phi_\mathbf{k}).
\end{equation} 
Here $A$ is a (constant) amplitude that is chosen so that the resulting {\it rms} velocity fluctuations are of order 
unity, $\langle u^2 \rangle \approx 1$. The $\mathbf{k}$-dependent amplitude, $a_{\mathbf{k}}$, is chosen with the aim of concentrating the 
energy in the velocity fluctuations at scales that are neither too large (since we wish to limit the effects of the 
boundary conditions) nor too small (since the magnetic field will grow on smaller scales and we must be 
able to resolve both and conduct the simulation over a number of eddy turnover times at the largest scale). We choose the Gaussian profile
\begin{eqnarray}
\label{eq:B}
a_{\mathbf k}&=&k^2 \left( \frac{3 y_0^5}{32 \pi^{3/2}}\right)^{1/2}\exp{\left(-\frac{k^2 y_0^2}{8}\right)},
\end{eqnarray}
where $k=|\mathbf{k}|$ and we take $y_0=\pi/4$. At each timestep the unit vector 
$\mathbf{\hat e_k}(t)$ is rotated about one of the three coordinate axes at random by an amount $\theta_\mathbf{k}(t)$, with 
$\theta_\mathbf{k}(t)$ 
and the random phase $\phi_\mathbf{k}(t)$ being drawn independently from uniform distributions on $[-\pi,\pi]$ at each 
timestep. 
This ensures that the force is short-time correlated. Its correlation time is of the order of the timestep ($\delta t$) of the numerical simulations. By decreasing the value 
of the viscosity we can then vary the correlation time of 
the flow from being approximately equal to that of the force to much larger values. This is an important point that 
we shall 
return to below. Homogeneity and isotropy of the force is maintained by rotating the vector $\mathbf{\hat 
e_k}$ around on the unit sphere. The force is guaranteed to be real in physical space (requiring  
$\mathbf{f}(-\mathbf{k},t)=\mathbf{f}^*(\mathbf{k},t)$) by initialising $\mathbf{\hat e_{\pm k}}$ appropriately and 
choosing equal seeds for the random number generators for the complex conjugate pairs. It can be shown that the 
spatial correlator of the force takes the form 
(\ref{eq:kappa_ij}) with $g^{force}(r) \equiv 0$ and 
$\kappa_L^{force}(r)=(4/y_0^4)(4r^4/y_0^4-28r^2/y_0^2+35)\exp(-r^2/y_0^2)$, which falls to zero beyond $r\approx \pi/2$ and 
hence satisfies the boundary conditions that the Kazantsev model assumes for the flow.

For chosen values of $\nu$ and $\eta$ (see below) equations (\ref{eq:induction},\ref{eq:momentum}) are solved using standard pseudospectral methods with a grid 
resolution of $256^3$ mesh points (for a detailed description of the numerical method see 
\cite{cattaneoew2003}). 
Initially equation (\ref{eq:momentum}) is evolved in isolation until the statistically steady state of the flow is 
reached, which is confirmed by observing 
the time evolution of the energy of the velocity fluctuations. A weak seed magnetic field is then introduced. After 
a short transient evolution the field settles onto the fastest growing eigenfunction and we measure the following 
quantities. 

First, it is necessary to compute the correlation time of the flow, $\tau_c$ say. The spatial correlators and spectra defined below represent time 
averages over a collection of statistically independent samples, i.e. samples separated by an interval of the order 
of the correlation time of the largest turbulent eddy. We define $\tau_c$ to be the interval over which the 
temporal correlator of the velocity 
falls to half of its original value, i.e. the $\tau$ for which
\begin{equation}
\label{eq:tau_c}
\langle u_i(\mathbf{x},t)u_i(\mathbf{x},t+\tau)\rangle_{t,V}=0.5\langle 
u_i(\mathbf{x},t)u_i(\mathbf{x},t)\rangle_{t,V}.
\end{equation}
Hereafter, angled brackets with subscript $t$ and/or $V$ 
denote averages over time and/or volume, respectively.
The quantities defined below are computed from data sets that consist of approximately 50-100 samples 
separated by an interval of length $\tau_c$. 

Second, we measure (twice) the exponential growth rate ($\lambda$) of the magnetic field by fitting the function 
$\exp(\lambda t)$ to $\langle \mathbf{B}^2 (\mathbf{x},t) \rangle_V$. We 
also measure the normalised magnetic spectrum 
\begin{equation}
\label{eq:EB}
\hat E_B(k)=2 \pi k^2 \langle \frac{ B_i^*(\mathbf{k},t) B_i(\mathbf{k}, t)} {\langle \mathbf{B}^2(\mathbf{x},t) 
\rangle_V }
\rangle_{t,\phi,\theta}
\end{equation}
where $*$ denotes the complex conjugate and $\theta$ and $\phi$ are the polar and azimuthal angles of the 
spherical coordinate 
system onto which the function of $\mathbf{k}$ is interpolated in order to construct the isotropic function of 
$k=|\mathbf{k}|$.

We now wish to compare the growth rate and the magnetic spectrum with that predicted by the Kazanstev model.  In 
order to solve the Kazanstev equation (\ref{eq:M_L}) we require the spatial correlator of the velocity 
$\kappa_L(r)$, which we compute as follows. First, we calculate the diffusivity spectrum from the 
numerical simulations. We take the trace of equation (\ref{eq:corr_v_k})
with $\mathbf{k}=\mathbf{k}'$ and integrate over $\tau$, thereby removing the $\delta$-function in time and yielding
\begin{eqnarray}
\label{eq:corr_v_int}
F(k)&=& \frac{1}{2}\langle u_i^*(\mathbf{k},t) \int_{-\infty}^{\infty} u_i(\mathbf{k},t+\tau)d\tau \rangle, \nonumber \\
&\approx& \frac{1}{2} \langle u_i^*(\mathbf{k},t) \int_{\bar t=t-n\tau_c}^{\bar 
t=t+n\tau_c} u_i(\mathbf{k}, \bar t)d \bar t \rangle_{t,\phi,\theta},
\end{eqnarray}
To obtain the final expression we have replaced the ensemble average with an average over time $t$ (assuming ergodicity), replaced the integral 
over 
$\tau \in [-\infty, \infty]$ by an integral over a finite interval that is sufficiently 
larger than the velocity correlation time $\tau_c$ in order to ensure convergence ($n$ is the smallest integer required for convergence), and integrated over the direction of $\mathbf{k}$. We then determine the spatial correlator $\kappa_L(r)$ from $F(k)$ by numerically integrating equation (\ref{eq:kappa_L_F}) using a modified Simpson's rule.\footnote{The integration is computed by using a piecewise parabolic interpolation for the non-harmonic functions of $k$ in equation (\ref{eq:kappa_L_F}). The result is then multiplied by the sine and cosine functions and integrated analytically.}

We then seek solutions of equation (\ref{eq:M_L}) in the form 
$M_L(r,t)=M_L(r)\exp(\lambda t)$. The equations are solved numerically using a fourth-order Runge-Kutta 
method (for details see \cite{malyshkinb2007}).  The magnetic spectrum function $F_B(k,t)$ can be determined from $M_L(r,t)$ by 
(see \cite{moniny1971}) 
\begin{equation}
\label{eq:M_L_F_B}
F_B(k,t)=\frac{1}{4 \pi^2 k} \int_{0}^{\infty} [r \sin(kr) -r^2 k \cos(kr)] M_L(r,t) dr
\end{equation}
In the next section we shall
compare the Kazantsev magnetic spectrum $2 \pi k^2F_B(k)$ and the growth rate $\lambda$ with the magnetic spectrum $\hat E_B(k)$ and the growth rate that are obtained from the direct numerical simulations.

\section{Results}

\begin{figure} [tbp]
\begin{center}
\includegraphics[scale=0.6]{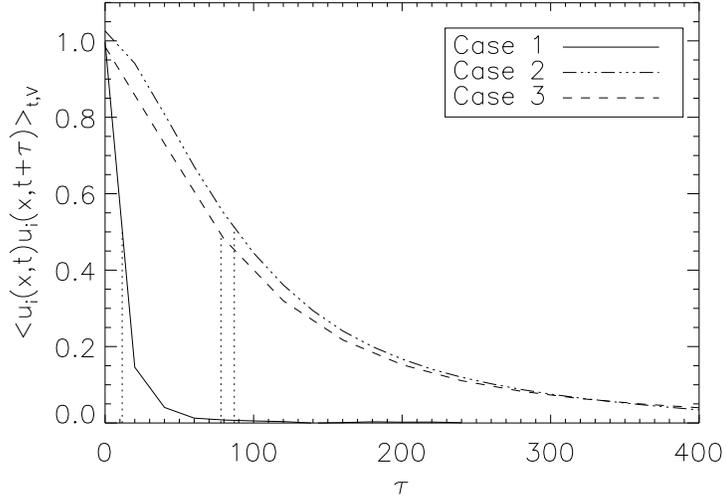}
\end{center}
\vskip2mm
\caption{The temporal correlation function $\langle u_i(\mathbf{x},t)u_i(\mathbf{x},t+\tau)\rangle_{t,V}$ for each of the three flows. The vertical dotted lines identify the correlation time, i.e. the time at which the correlator has decreased to half of its value at $\tau=0$ (see equation~(\ref{eq:tau_c})).}
\label{fig:correlation_time}
\end{figure}

\begin{figure} [tbp]
\begin{center}
\includegraphics[scale=0.4]{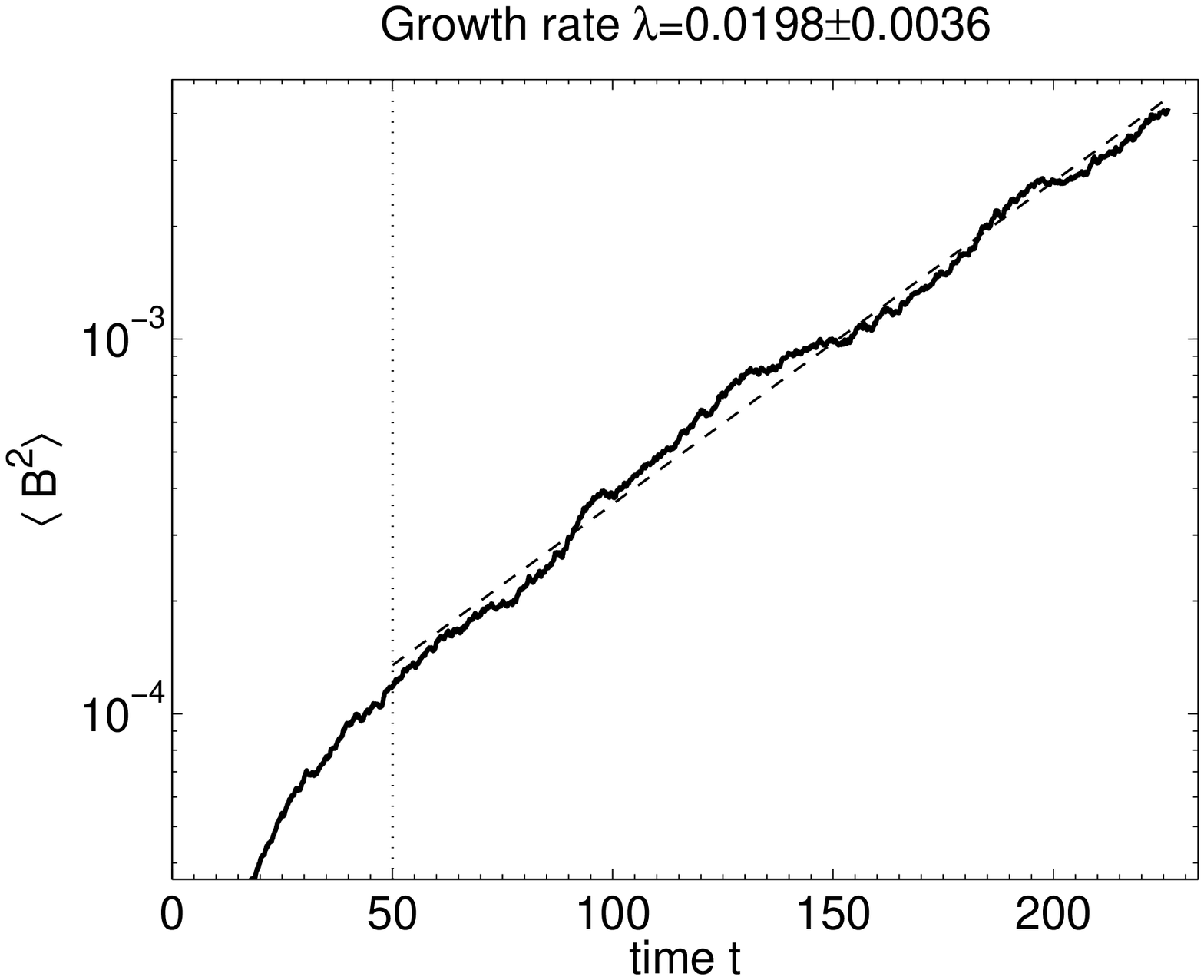}
\includegraphics[scale=0.4]{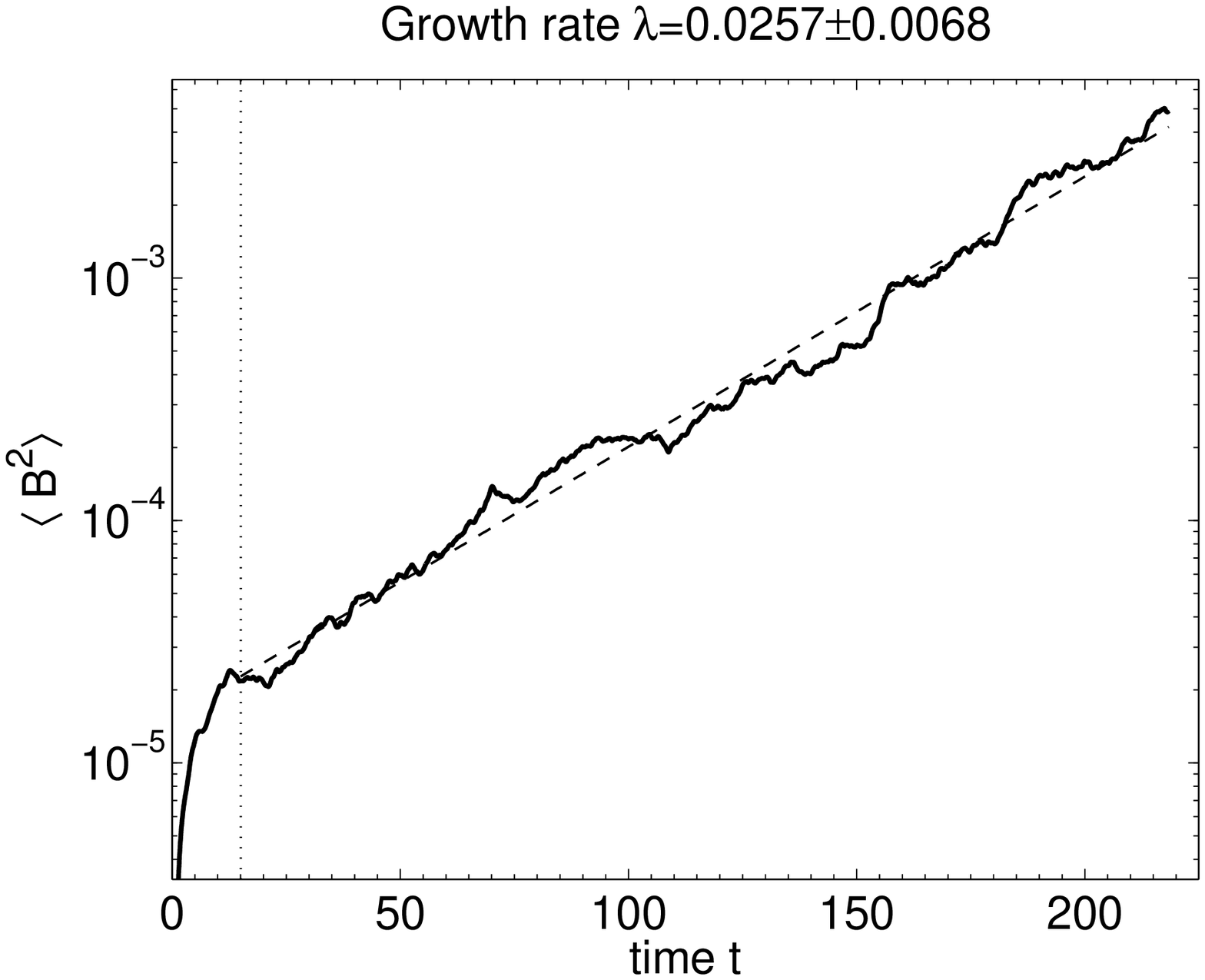}
\includegraphics[scale=0.4]{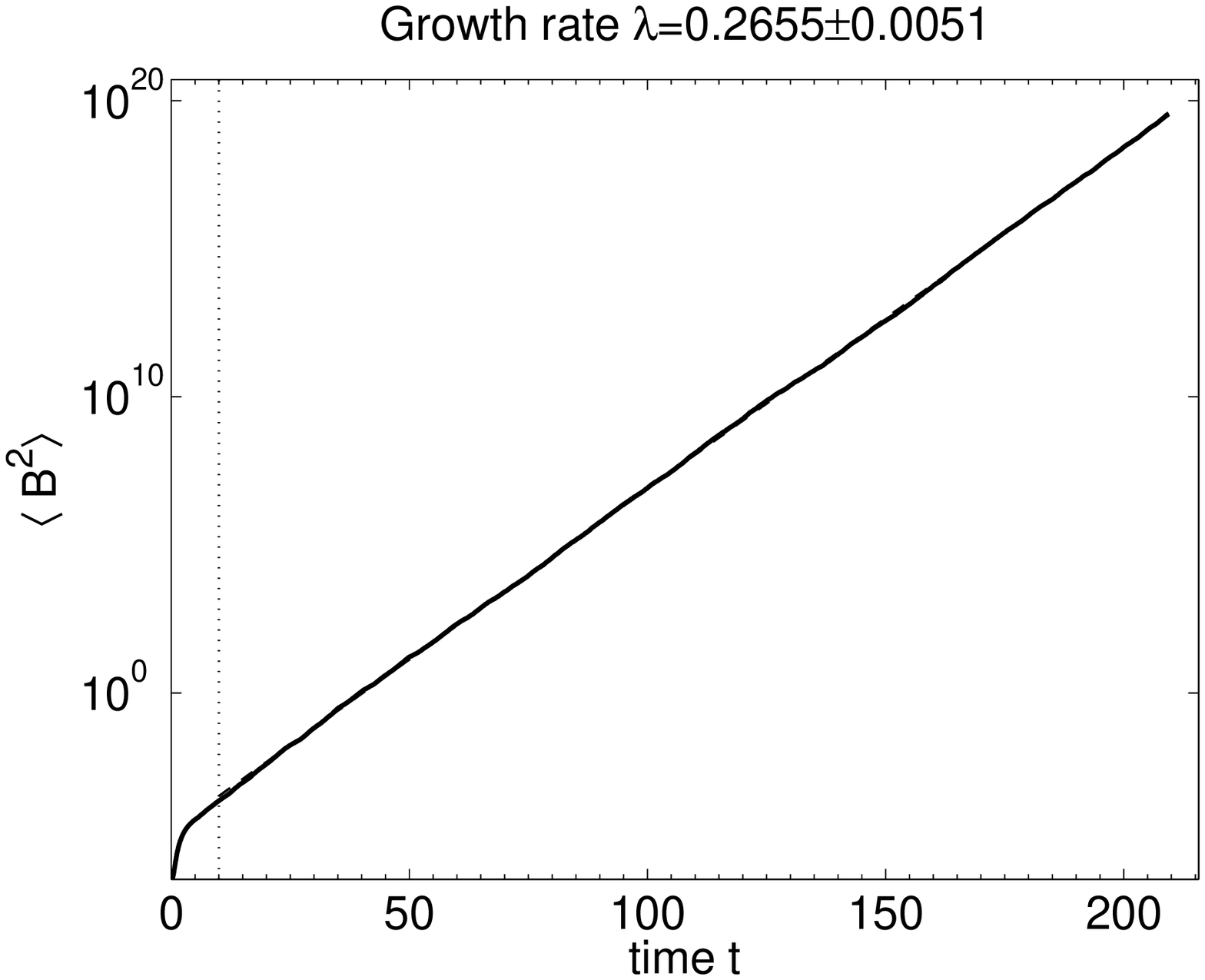}
\end{center}
\vskip2mm
\caption{The magnetic energy as a function of time from the direct numerical simulations: (a) Case 1, (b) Case 2, (c) Case 3. The dotted lines mark the beginning of the time interval over which the growth rate is measured (see the text).}
\label{fig:growth_rate}
\end{figure}

\begin{figure} [tbp]
\begin{center}
\includegraphics[scale=0.4]{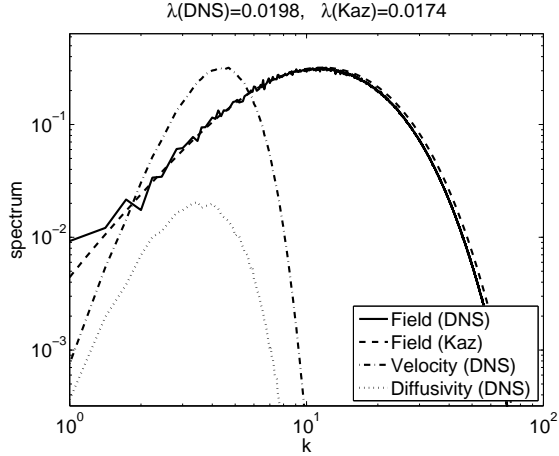}
\end{center}
\vskip2mm
\caption{A comparison of the magnetic spectrum from the direct numerical simulations (solid line) and the Kazanstev model (dashed line) for Case 1. For purposes of reference, the dotted line illustrates the diffusivity spectrum $2 \pi k^2 F(k)$ 
($n=12$) and the dashed-dotted line illustrates the velocity spectrum (defined analogously to the magnetic spectrum, see equation (\ref{eq:EB})).}
\label{fig:mag_spec_1}
\end{figure}
\begin{figure} [tbp]
\begin{center}
\includegraphics[scale=0.4]{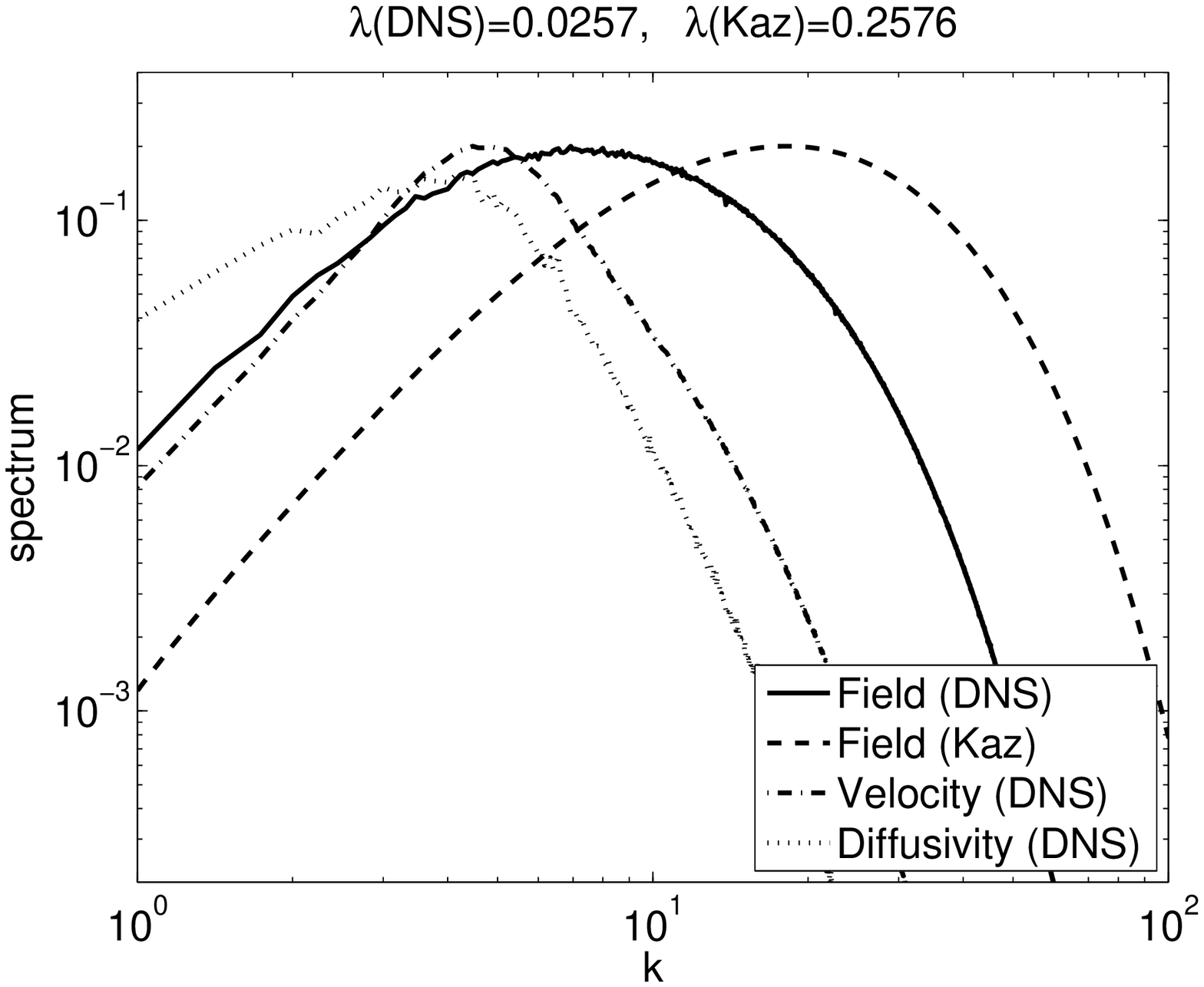}
\includegraphics[scale=0.4]{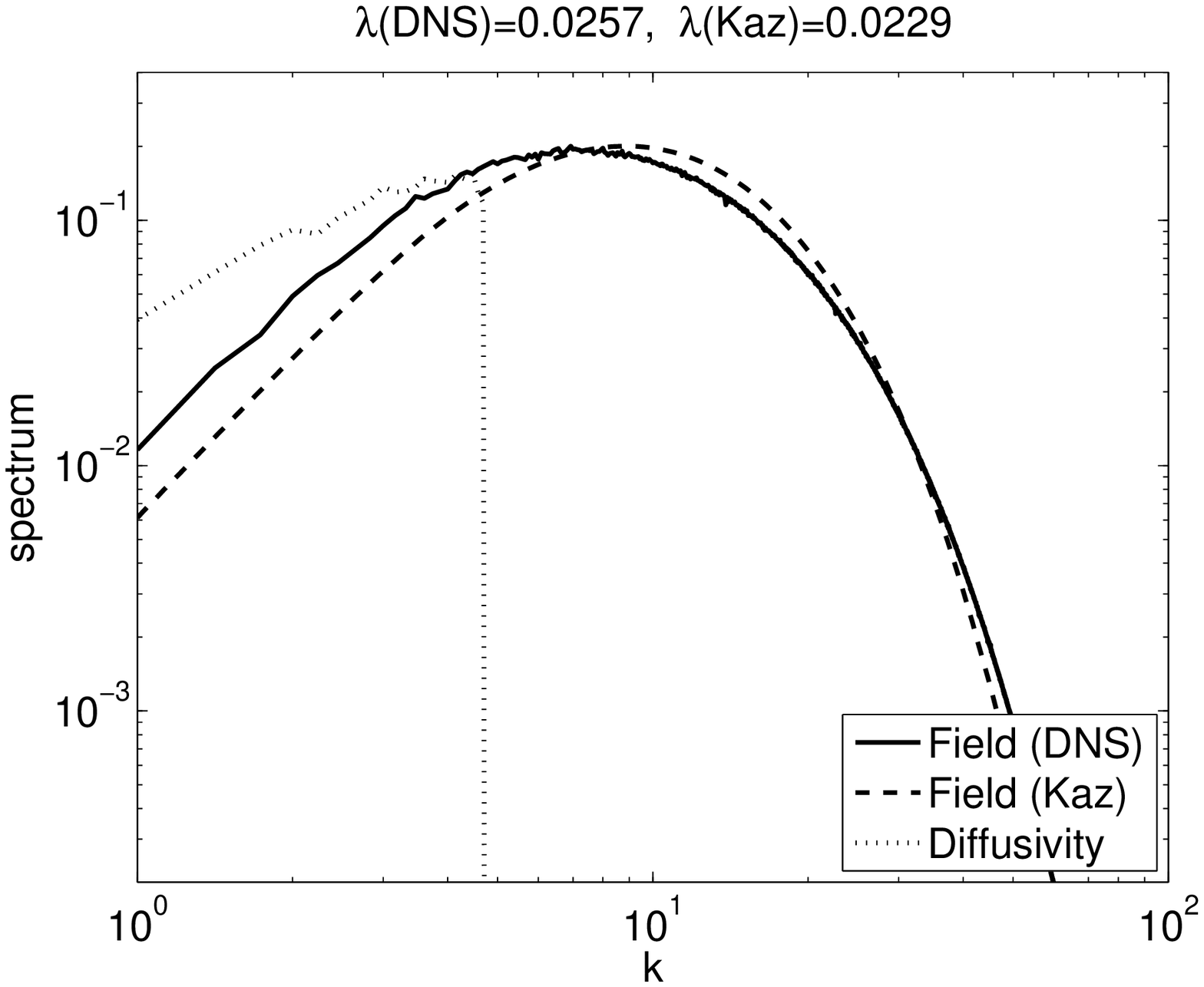}
\end{center}
\vskip2mm
\caption{Left: As for Figure~\ref{fig:mag_spec_1} except for Case 2. Note that the Kazantsev magnetic spectrum is shifted to the right by a factor of approximately 2.6. Right: Filtering the diffusivity spectrum at $k_f=4.85$ brings the growth rates into agreement (within the error bounds) and simultaneously aligns the magnetic spectra.}
\label{fig:mag_spec_2}
\end{figure}
\begin{figure} [tbp]
\begin{center}
\includegraphics[scale=0.4]{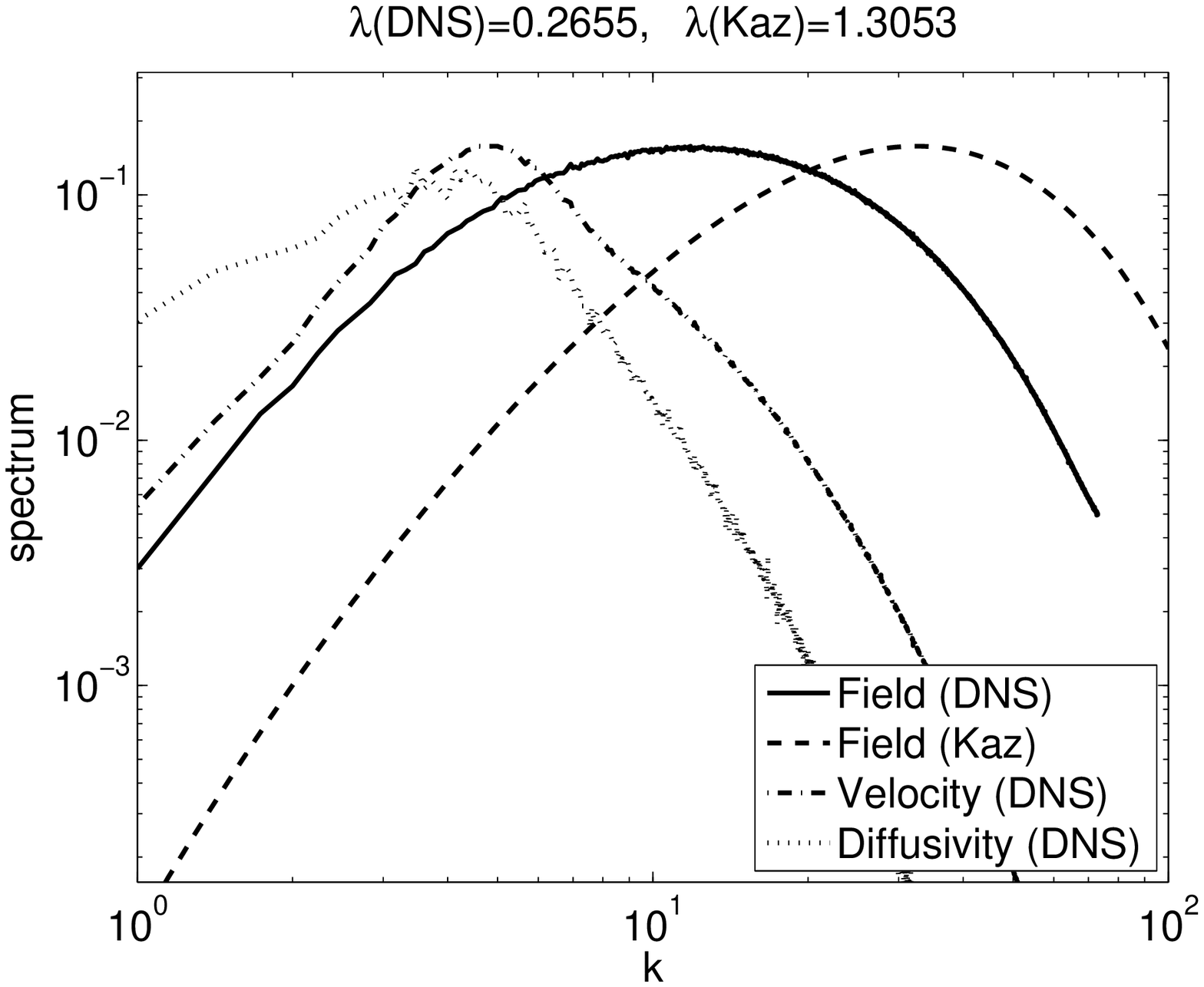}
\includegraphics[scale=0.4]{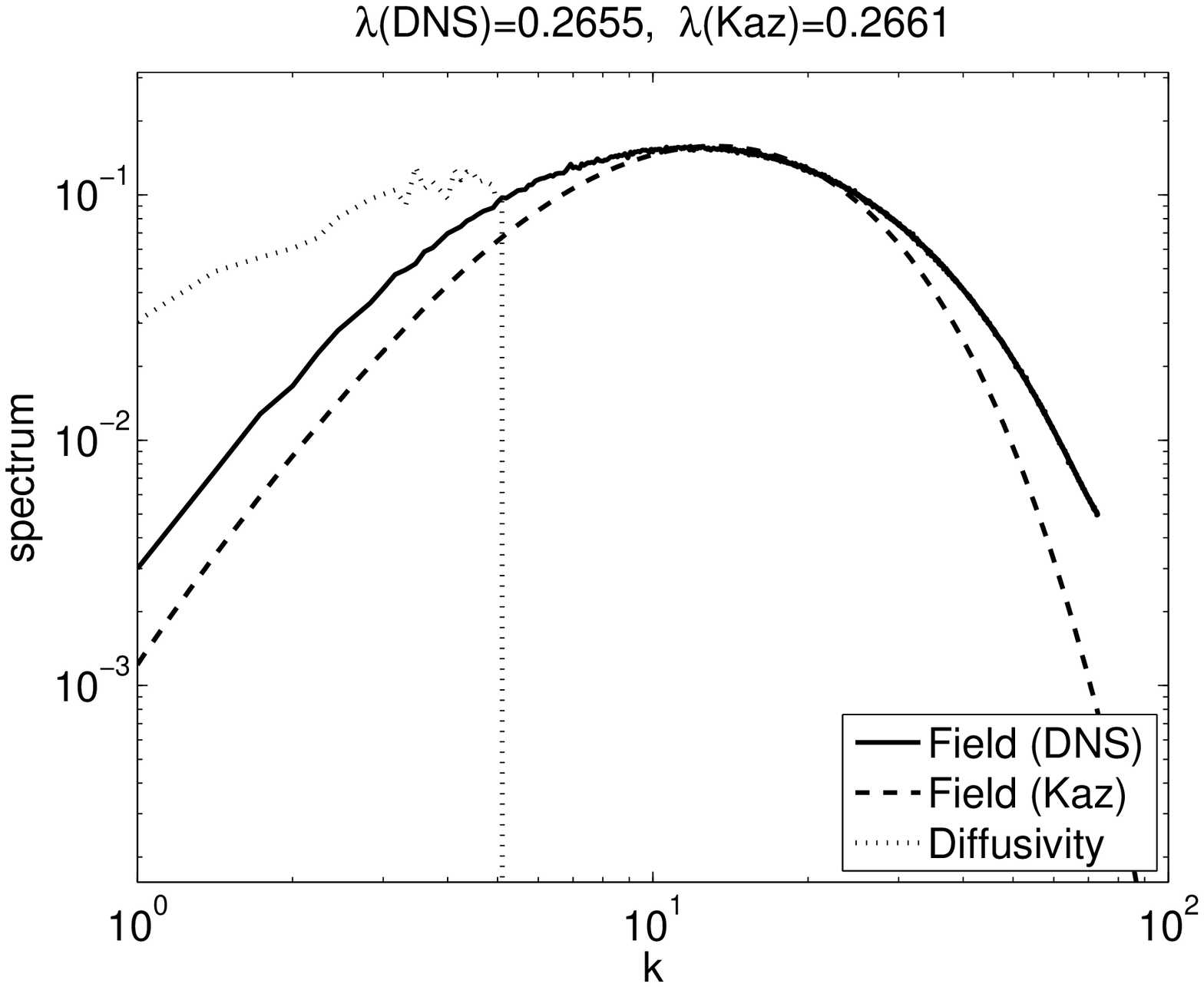}
\end{center}
\vskip2mm
\caption{As for Figure~\ref{fig:mag_spec_2} except for Case 3. Compared to the direct numerical simulation, the Kazantsev spectrum is shifted to the right by a factor of approximately 2.7. Filtering the diffusivity spectrum at $k_f=5.10$ brings the growth rates and magnetic spectra into approximate agreement.}
\label{fig:mag_spec_3}
\end{figure}

First, we concentrate on the direct numerical simulations (DNS). We report the results of three cases. Case 1 represents a 
strongly diffusive test case. We take $\nu=1$ in anticipation that the velocity 
will then inherit the properties of the force, especially its short-time correlation. Hence we expect that the results of the numerical simulations and the Kazantsev model will be in agreement. We choose a small value of 
the magnetic resistivity, $\eta=0.0008$, ensuring that the dynamo is excited and the magnetic field is resolved. Case 2 represents a 
more turbulent scenario. We take $\nu=\eta=0.007$. Thus the nonlinear term in the 
momentum equation (\ref{eq:momentum}) is important and the turbulent eddies should have a correlation time 
$\tau_c$ that is of the order of the eddy turnover time and is much longer than the correlation time of the force. This particular value of $\eta$ ($\nu$) is chosen to correspond to the case of a marginal dynamo, while ensuring that both the velocity and the magnetic field are resolved. Case 3 has 
$\nu=\eta=0.0025$, representing a situation similar to that of Case 2 (with the flow being turbulent and the ratio 
$\nu/\eta=1$) except that we expect that the magnetic growth rate will be larger (see, e.g., \cite{schekochihinetal2004}).

In all three cases, $u_{rms}=\langle |\mathbf{u}|^2\rangle^{1/2} \approx 1.0$ and $\delta t \approx 0.0055$. A physical 
(eddy turnover) time unit $\tau_p=y_0/u_{rms} \approx 0.8$ therefore corresponds to a time interval of 
approximately $145 \delta t$. This is to be compared with the correlation time of each the flows, $\tau_j$, say, where the numeric subscript $j=1,2,3$ will henceforth denote the case number. Figure~{\ref{fig:correlation_time} illustrates that $\tau_1 \approx 
10 \delta t$,  $\tau_2 \approx 90 \delta t$ and $\tau_3 \approx 80 \delta t$.
Figure~\ref{fig:growth_rate} shows the time series of the magnetic 
energy for each case. The growth rates are estimated to be $\lambda_1 \approx 0.0198 \pm 0.0036$, $\lambda_2 
\approx 0.0257 \pm 0.0068$ and $\lambda_3 \approx 0.2655 \pm 0.0051$. The error bars are estimated by the 
bootstrap method, as follows. From the time interval $T$ over which 
the growth rate of the field is measured (i.e.~to the right of the vertical 
dotted lines in Figure~\ref{fig:growth_rate}) 
we randomly choose $I=1/\Delta t$ sub-intervals 
of length $\Delta t$, where $\Delta t$ is fixed and chosen such that $\delta t \ll  \Delta t \ll T$. We estimate the field growth rates $\lambda_i$ 
over each of the sub-intervals, $i=1,2,...,I$. The standard deviation 
of the mean rate ${\bar\lambda}=(1/I)\sum\lambda_i$ is then
estimated as $std(\bar\lambda)=std(\lambda_i)/\sqrt{I}$. We then repeat the procedure a number of times with different $\Delta t$ and compute the average $\langle std(\bar\lambda)\rangle$. It is noteworthy that $std(\bar\lambda)$ is not very sensitive to the value of $\Delta t$ provided that $\delta t \ll  \Delta t \ll T$. The error bars are set to ${}\pm 1.96 \langle std(\bar\lambda)\rangle$, which corresponds to the $95\%$ confidence 
interval (assuming asymptotic normality). 

The solid line in Figure~\ref{fig:mag_spec_1} illustrates the magnetic spectra $\hat E_B(k)$ obtained from the numerical simulations for Case 1. The left-hand panels of Figure~\ref{fig:mag_spec_2} and Figure~\ref{fig:mag_spec_3} show the corresponding results for Cases 2 and 3, respectively. We note that the magnetic energy peaks at $k_1^*\approx 11$, $k_2^*\approx 7$ and $k_3^*\approx 12$. The dotted lines denote the converged diffusivity spectra, $2 \pi k^2F(k)$, calculated by using equation (\ref{eq:corr_v_int}). The velocity energy spectra (defined analogously to $\hat E_B(k)$) are shown by the dashed-dotted lines. 

We now turn to the solutions of the Kazantsev model. For each of the three cases, using relation (\ref{eq:kappa_L_F}) to calculate $\kappa_L(r)$ from the diffusivity spectrum, we solve equation (\ref{eq:M_L}) for the fastest growing eigenmode. We find $\lambda_{1}=0.0174$, $\lambda_{2}=0.2576$ and $\lambda_{3}=1.3053$. The Kazantsev magnetic spectra are shown by the dashed lines in Figure~\ref{fig:mag_spec_1} (Case 1) and the left-hand panels of Figure \ref{fig:mag_spec_2} (Case 2) and Figure \ref{fig:mag_spec_3} (Case 3).

As expected, the agreement between the numerical simulations and Kazantsev results is good for Case 1. 
The growth rates agree within the error bounds and the spectra peak at approximately the same wavenumber. 
However, much less can be said for cases 2 and 3. The growth rates disagree by factors of approximately $11$ 
(Case 2) and $5$ (Case 3) and the peak in the Kazantsev magnetic energy spectra are shifted 
towards larger wavenumbers (smaller scales) by factors of approximately 2.6 (Case 2) and 2.7 (Case 3). What is striking 
however is that the shape of the spectra are similar.  
Indeed, it appears as though the Kazantsev spectrum could be matched to that of the numerical simulations by a simple translation. 

In an attempt to determine a possible reason for the apparent translation, we proceed to investigate the sensitivity 
of the Kazantsev dynamo to the energy containing scales of the velocity field. We do this by applying a spectral 
filter to the DNS velocity before we supply it to the Kazantsev model. The 
technique was recently used by \cite{tobiasc2008b} to investigate the role of coherent structures on kinematic 
dynamo action. Here we apply a Heaviside filter to the diffusivity spectrum $2\pi k^2F(k)$ such that all modes with $k>k_f$ are 
set to zero. We then assess the changes that occur to the Kazantsev dynamo growth rate and the magnetic spectrum 
as $k_f$ is decreased.  

It is not surprising that applying the spectral filter at large wavenumbers (i.e.~removing the energetically insignificant velocity fluctuations) has little effect 
on the dynamo, however, it is interesting that the growth rate can be made to pass through the value obtained in the numerical simulations by decreasing the filtering scale $k_f$. In particular, we find that filtering at $k_f=4.85$ (Case 2) and $k_f=5.10$ (Case 3) yields the Kazantsev growth rates $\lambda_{2_{f}}=0.0229$ (Case 2) and $\lambda_{3_{f}}=0.2661$ (Case 3), i.e. brings them into agreement (within the error bounds) with those of the DNS. We note that in both cases the optimal filtering scale is very near to the peak energy containing scale $k_2^*=4.5$ (Case 2) and $k_3^*=5$. What is even more striking is that for the optimal filtering scale the corresponding magnetic spectra are translated to the left so that they almost lay on top of those obtained from the DNS. The results are shown in the right-hand panels of Figure~\ref{fig:mag_spec_2} (Case 2) and Figure~\ref{fig:mag_spec_3} (Case 3). Thus the filtering procedure simultaneously brings both the Kazantsev growth rate and magnetic spectrum into agreement with those obtained from the direct numerical simulations.  We note that this result is not specific to the cases detailed here. We have also checked that the procedure produces similar results when, for example,  the flow is turbulent and the ratio $\eta/\nu$ is slightly larger than unity. The complementary case of small $\eta/\nu$ is also relevant astrophysically but it is inaccessible with the available computational power. We have also checked that the results are not specific to our choice of the forcing function, with similar results being obtained if the amplitude of the force $a_{\mathbf k}$ is replaced by $a_{\mathbf k}/k^2$.

\section{Discussion}

In the Kazantsev dynamo model the statistical properties of the flow are given and one seeks the 
corresponding statistical properties of the generated magnetic field. The Kazantsev flow is 
instantaneously correlated in time, Gaussian distributed with zero mean and homogeneous and isotropic in space. 
In the nonhelical case it is then possible to reduce the dynamo problem to solving a single ordinary differential equation for 
the longitudinal correlation function of the magnetic field. This is to be compared with solving the deterministic 
system, in which a specific flow is prescribed and one solves a partial differential equation for the spatial and 
temporal evolution of the magnetic field. The Kazantsev model therefore allows a considerable saving in 
computational effort and permits the exploration of a parameter regime that is much more extreme than would be 
possible otherwise. For these reasons the Kazantsev model has become a popular tool for studying 
astrophysical dynamo action. 

It is necessary however to address the issue of whether astrophysical flows are likely to satisfy the model assumptions, i.e. to determine whether a $\delta$-correlated Gaussian random flow is a suitable model of the inertial range of 
astrophysical turbulence. If it is not, we must then determine whether the results of the Kazantsev model depend sensitively 
on the assumed statistical properties. In this paper we have concentrated on the issue of temporal correlations in the flow, as it is believed that finite time correlations will naturally arise through the equations of electrically conducting fluid dynamics. We have not studied the effects of relaxing the assumptions made regarding Gaussianity, homogeneity and 
isotropy in space and we have also restricted our attention to non-helical flows. We have investigated the 
kinematic dynamo properties of flows with similar spatial structure but different degrees of temporal 
correlations. We have found that while the Kazantsev model accurately describes the system when the flow is 
short time correlated, increasing the correlation time results in large differences between the Kazantsev prediction 
and the evolution of the MHD system, both in terms of the dynamo growth rate and the magnetic energy 
spectrum. In particular, it appears that the Kazantsev model overestimates the growth rate and yields a magnetic 
spectrum that is peaked at smaller scales (larger $k$).

We believe that we have traced the source of the problem in the Kazantsev model to a mistreatment of the small, sub-viscous scales of the velocity field. It appears that the model overestimates stretching or underestimates magnetic diffusion at small scales. In particular, we have shown that the Kazantsev model's magnetic growth rate and the peak in the magnetic spectrum can be simultaneously brought into agreement with the results of the DNS by filtering the diffusivity spectrum at scales larger than the scale at which the velocity spectrum peaks. We believe that this is an important non-trivial result. 

At this stage we cannot offer a full theoretical explanation of our results. However, we would like to suggest the following qualitative physical picture for why the Kazantsev model mistreats the small sub-viscous scales and why the filtering procedure works. We note that the Kazantsev model assumes that magnetic field is amplified by turbulent eddies that are $\delta$-correlated in time at all scales. In contrast, in real turbulence the small-scale eddies are not short-time correlated, they rotate together with the viscous eddy and have scale-independent correlation times that are approximately equal to that of the viscous eddy. As a result, the sub-viscous eddies do not act independently to stretch and fold the magnetic field lines as efficiently as the Kazantsev model assumes. In the direct numerical simulations presented here the magnetic Prandtl number is chosen to be unity and the magnetic energy peaks at the sub-viscous scales (see Figures~\ref{fig:mag_spec_2} and \ref{fig:mag_spec_3}) at which smooth velocity fluctuations are poorly modelled as a short-time-correlated Gaussian random process. By filtering out the incorrectly treated velocity fluctuations at small scales we do indeed achieve a better agreement. In addition, it is possible that even better agreement with the Kazantsev model could be achieved if the inertial range of our direct numerical simulations could be extended (the extremely short inertial range that we currently have is a result of the requirement of limiting the effects of the periodic boundary conditions together with the constraints imposed by presently available computational power). This issue is certainly worthy of future investigation. 

We would also like to note that a number of other ways of artificially modifying either of the two inputs to the Kazantsev model (i.e. $\eta$ and $\kappa_L(r)$) were tried but none of them were successful in the sense of simultaneously aligning both the growth rates and the magnetic spectra. For example, we have found that the growth rates alone can be aligned in a number of ways, for example by artificially `squeezing' $F(k)$ (which is equivalent to rescaling the wavenumber $k \rightarrow ak$, where $a$ is a constant), `squashing' $F(k)$ (i.e. $F(k) \rightarrow aF(k)$, where $a$ is a constant), or by simply retaining the diffusivity spectrum from the numerical simulations but increasing the magnetic diffusivity $\eta$ in the Kazantsev model. However, in all cases the shift in the magnetic spectrum is very slight and it does not align with that from the DNS. Filtering the diffusivity spectrum at large scales (small $k$) also didn't work. 

One possible reason why it is difficult to match the Kazantsev predictions with the results of the DNS is that, due to numerical constraints, we have a magnetic Prandtl number $P_m=\nu/\eta \approx 1$ in our simulations. In the small magnetic Prandtl number regime, $P_m<<1$, the Kazantsev model may better describe dynamo action since in this regime the magnetic field is generated by a rough velocity field that can be treated as a turbulent diffusion process. Indeed, within the inertial range of a Kolmogorov flow the correlation time $\tau_c \sim l^2/D_l$ (where $D_l\sim v_ll$ is the turbulent diffusivity) can be estimated as $\tau_c \sim l^2/(l^{1/3}l)\sim l^{2/3}$ and is comparable to the turnover time $l/v_l$. In the opposite limit, $P_m>>1$, the velocity fluctuations are smooth on the resistive scales and thus $\kappa(r)=\kappa_L(0)-\kappa_L(r)+2\eta \approx -(1/2)\kappa_L''(0)r^2+2\eta$ in equation (\ref{eq:M_L}). Thus for a given magnetic diffusivity, the Kazantzev model is essentially dependent only on the single parameter $\kappa_L''(0)$. Even though in this case the sub-viscous eddies are not short-time correlated, and hence by the above arguments the Kazantsev model is not expected to work perfectly, essentially any change to the diffusivity spectrum that adjusts $\kappa_L''(0)$ to the right value in the sense of matching the magnetic growth rate must also simultaneously align the magnetic spectra. Unfortunately, numerical tests of the small and large magnetic Prandtl number regimes must await further increase in computational power.

\acknowledgments
This work was supported by the NSF Center for Magnetic
Self-Organization in Laboratory and Astrophysical Plasmas
at the University of Chicago and the University of Wisconsin - Madison, the US DoE awards DE-FG02-07ER54932, DE-SC0003888, DE-SC0001794, and the NSF grant PHY-0903872. This research used resources of the 
Argonne Leadership Computing Facility at Argonne National Laboratory, which is supported by the Office of 
Science of the U.S. Department of Energy under contract DE-AC02-06CH11357.

\end{document}